\documentclass[letterpaper, 10 pt, conference]{ieeeconf}

\IEEEoverridecommandlockouts                      

\usepackage{cite}
\usepackage{amsmath,amssymb,amsfonts}
\usepackage{algorithm}
\usepackage{algpseudocode}
\usepackage{graphicx}
\usepackage{subcaption}
\usepackage{textcomp}
\usepackage{xcolor}
\usepackage{mathtools}
\usepackage{hyperref}
\newtheorem{assumption}{Assumption}
\def\BibTeX{{\rm B\kern-.05em{\sc i\kern-.025em b}\kern-.08em
    T\kern-.1667em\lower.7ex\hbox{E}\kern-.125emX}}
\algnewcommand{\StepComment}[1]{\Statex \(\triangleright\)~#1}

\title{\LARGE \bf Distributed Model-Free Turbine Repositioning for Wake Overlap Minimization in Floating Offshore Wind Farms$^{*}$}

\author{Zekai Chen$^{1}$ and Ryozo Nagamune$^{1}$% <-this % stops a space
\thanks{*This work was supported by the Natural Sciences and Engineering
Research Council of Canada (NSERC) and University of British Columbia Four Year Doctoral Fellowship (4YF).}% <-this % stops a space
\thanks{$^{1}$Zekai Chen and Ryozo Nagamune are with the Department of Mechanical
Engineering, the University of British Columbia, Vancouver, BC V6T 1Z4,
Canada {\tt\small zkaichen@student.ubc.ca, nagamune@mech.ubc.ca}}%
}

\begin{document}

\maketitle
\thispagestyle{empty}
\pagestyle{empty}

%%%%%%%%%%%%%%%%%%%%%%%%%%%%%%%%%%%%%%%%%%%%%%%%%%%%%%%%%%%%%%%%%%%%%%%%%%%%%%%%
\begin{abstract}
This paper presents a model-free method for turbine repositioning control in floating offshore wind farms. The method uses a distributed online optimization framework to mitigate the wake effect. Conventional approaches rely on centralized, model-based control architectures, which suffer from poor practical feasibility. To overcome this limitation and address nonconvexity and inaccessible gradient information, we incorporate physics-informed knowledge of wake dynamics into the control system design. We demonstrate the effectiveness of the proposed method by comparing it with a centralized, model-based baseline in mid-fidelity simulations. Results indicate that the proposed method achieves slightly better performance while significantly reducing computational cost.
\end{abstract}

% \textbf{Keywords:}
% Wind Farm Control

%%%%%%%%%%%%%%%%%%%%%%%%%%%%%%%%%%%%%%%%%%%%%%%%%%%%%%%%%%%%%%%%%%%%%%%%%%%%%%%%
\section{INTRODUCTION} \label{sec:intro}
Floating offshore wind energy is a highly promising clean energy source. Unlike fixed-bottom offshore wind farms, which are limited to water depths up to about 60 meters, floating offshore wind farms (FOWFs) can be installed in deeper water, unlocking access to abundant wind energy resources~\cite{niu2024floating}. Despite substantial potential, mass commercialization of FOWFs is hindered by the high levelized cost of energy. One primary factor is the wake effect: when turbines are placed in an array, upstream turbines extract energy from the incoming airflow, creating regions of air with reduced wind speed and higher turbulence intensity behind them. Downstream turbines within these regions suffer from reduced power capture and increased structural loads~\cite{burton2011wind}. For wind farms, the wake effect can result in annual revenue losses ranging from $20\%$ to $30\%$~\cite{busby2012wind}. 

To mitigate the wake effect, wind farm control has been developed to shift operation from ``greedy'' to ``cooperative''. In greedy operation, each turbine optimizes its own performance without accounting for the wake effect, whereas in cooperative operation, a controller coordinates all turbines to mitigate the wake effect and improve aggregate farm-level performance~\cite{WFFC_Review, kheirabadi2019quantitative}. For floating offshore wind turbines (FOWTs), the additional degrees of freedom enable a strategy called turbine repositioning~\cite{niu2024floating}. Using nacelle yaw \(\gamma\), this strategy actively adjusts the positions of FOWTs to reduce the overlap between the incoming wake and turbine rotor disks, thereby mitigating the wake effect.

Several previous studies have investigated methods to derive the optimal nacelle yaw angle. Studies of~\cite{kheirabadi2019modeling, kheirabadi2020real} address this by solving a nonlinear optimization problem based on a wind farm wake model with the goal of power maximization. A similar formulation has been adopted in multi-objective settings that also address load minimization~\cite{wei2025multi, niu2025power, huang2025bi} and power tracking~\cite{jard2025real}. Mei et al.~\cite{mei2025improving} further extend previous work by employing reinforcement learning to handle nonlinearity. Despite these contributions, these studies rely on low-fidelity, steady-state wake models with simplified assumptions, making the resulting controllers struggle to handle real-world uncertainty and unmodeled dynamics~\cite{boersma2017tutorial}. Additionally, their centralized control architectures suffer from intractable computational scaling as wind farm size grows, limiting practical deployment~\cite{kheirabadi2019quantitative}.

To address these limitations, we propose a model-free method based on the distributed online optimization framework~\cite{li2023survey}. To handle nonconvexity and inaccessible gradient information~\cite{marden2013model}, our method derives optimal nacelle yaw angles using local control logic informed by wake dynamics~\cite{kheirabadi2019quantitative}. Instead of explicit wake models, each FOWT uses local sensing signals to estimate wake dynamics~\cite{bottasso2018local, bertele2021wind}, and then the controller minimizes the cost function that is designed based on the objective of wake mitigation. This approach significantly reduces computational effort compared to centralized model-based methods. To our knowledge, this is the first model-free, distributed framework for FOWF turbine repositioning.

The contributions of this paper are as follows: (1) We propose a model-free, distributed method for turbine repositioning control that is computationally efficient. (2) We quantify the wake overlap using indirect wake dynamics detection based on each FOWT's available local sensing signals, thereby linking the objective of mitigating wake effects directly to the wake dynamics. (3) We validate the proposed method in FAST.Farm~\cite{FastFarm}.

The remainder of this paper is organized as follows. Section~\ref{sec:problem} presents the necessary definitions and assumptions and formulates the control problem. Section~\ref{sec:method} introduces the proposed framework, which is the main contribution of this work. Section~\ref{sec:case} validates the proposed method through a case study. Finally, Section~\ref{sec:conclusion} concludes the paper.
\section{Turbine Repositioning PROBLEM Formulation} \label{sec:problem}
In this section, we introduce the necessary assumptions and then formulate the turbine repositioning problem.

\subsection{Wind Farm Setup}
\begin{figure}[h!]
    \centering
    \includegraphics[width=.8\linewidth]{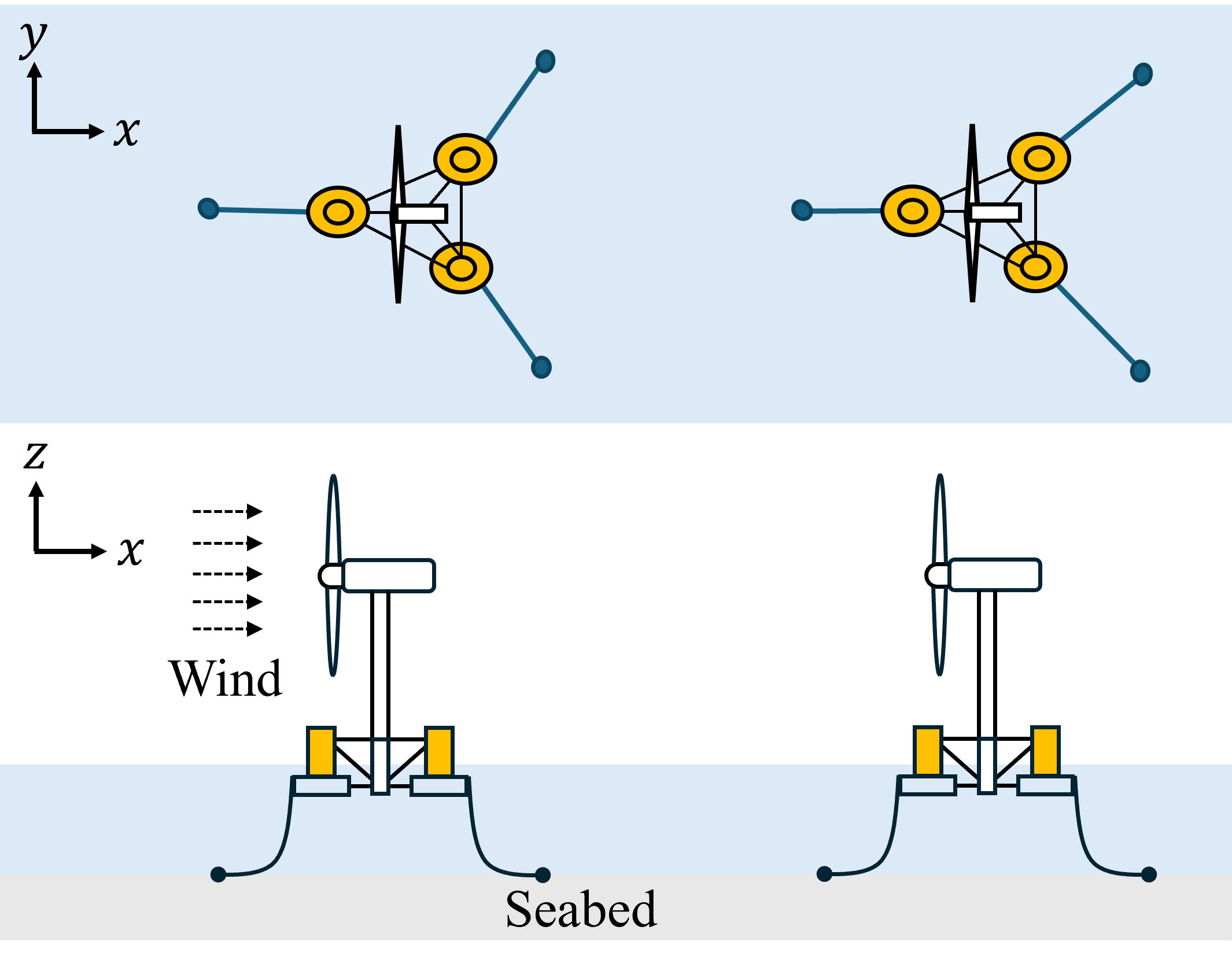}
    \caption{An array FOWF with \(2\) turbines. The top and bottom panels show the bird's-eye view and the side view.}
    \label{fig:fowf}
\end{figure}

We consider an array wind farm consisting of \(N\) FOWTs. Figure~\ref{fig:fowf} illustrates an example FOWF with~\(N=2\). Let \(n \in \{1,2,\dots, N\}\) index all FOWTs, ordered in the direction of wake propagation, so that FOWT\(_1\) is the most upstream turbine and FOWT\(_{n+1}\) is directly downstream of FOWT\(_n\). Let \(t\) denote the discrete time step, with \(T\) denoting the total number of time steps. Each FOWT\(_n\) is characterized by its rotor diameter \(D\) [m] and its nacelle yaw angle \(\gamma_{n,t}\) [deg] at time step \(t\). The inflow wind exerts force on the rotor of each FOWT\(_n\), producing the out-of-plane blade root bending load~\(M_{n,t} := \{M^{(1)}_{n,t}, M^{(2)}_{n,t}, M^{(3)}_{n,t}\}\) [MNm] for three different blades~\cite{bottasso2018local}. 

% \subsection{Notations}
% Let \(n \in [N]\) index all FOWTs, with \([N] \coloneqq \{1,2,\dots,N\}\), and let \(t\) denote the discrete time step. The cost of the entire FOWF at time \(t\) is denoted by \(f_t\), while \(f_{n,t}\) denotes the local cost of FOWT\(_n\) at \(t\). The nacelle yaw angle of FOWT\(_n\) at time \(t\) is denoted by \(\gamma_{n,t}\). Moreover, we define the neighbor of FOWT\(_n\) as its immediate upstream turbine FOWT\(_{n-1}\) for \(n \geq 2\), consistent with the wake propagation sequence. Consequently, the most upstream turbine FOWT\(_1\) has no neighbor. The azimuth angle and out-of-plane bending moment of blade \(j\) on a FOWT at \(t\) are denoted by \(\psi_{j,t}\) and \(M_{\mathrm{oop},j,t}\), respectively, with \(\psi_j = 0^\circ\) corresponding to the vertically upright position. In addition, \(D\) denotes the rotor diameter of a FOWT. Finally, $\text{sgn}(\cdot)$ denotes the operation to get the sign of a signal.

\subsection{Assumptions on Wind Farms}
In this work, we make the following assumptions. First, regarding baseline turbine-level control:
\begin{assumption} \label{assump:baseline}
    Each FOWT uses a baseline controller to regulate its stability and power output performance~\cite{NREL5MW}.
\end{assumption}

Based on Assumption~\ref{assump:baseline}, changing \(\gamma_{n,t}\) will not induce instability. Furthermore, we assume that:
\begin{assumption} \label{assump:freestream}
    The freestream turbine FOWT\(_1\) always operates in a greedy manner.
\end{assumption}

Under this assumption, FOWT\(_1\) always pursues the best attainable performance under the given wind condition. Moreover, we assume the following measurement availability to ensure practical feasibility~\cite{WFFC_Review}:
\begin{assumption} \label{assump:measurements}
    Sensors on each FOWT can measure its blade load \(M_{n,t}\) and yaw angle \(\gamma_{n,t}\) at every time step \(t\).
\end{assumption}

To enable distributed control, we allow FOWTs to communicate with one another. We define the neighbor of FOWT\(_n\) as the turbine that aerodynamically influences it. For simplicity, this neighbor is the immediate upstream turbine FOWT\(_{n-1}\). We then assume the following:
\begin{assumption} \label{assump:communication}
    FOWT\(_n\) can communicate with its neighboring turbine FOWT\(_{n-1}\) by exchanging \(M_{n,t}\) and \(\gamma_{n,t}\) at each time step \(t\). In addition, \(M_{1,t}\) from the freestream turbine FOWT\(_1\) is broadcast to all FOWTs.
\end{assumption}

Since FOWT\(_1\) represents the best attainable performance according to Assumption~\ref{assump:freestream}, we broadcast the freestream measurement \(M_{1,t}\) to all FOWTs so that it serves as a performance reference for all downstream turbines. 

\begin{figure}[h!]
    \centering
    \includegraphics[width=.8\linewidth]{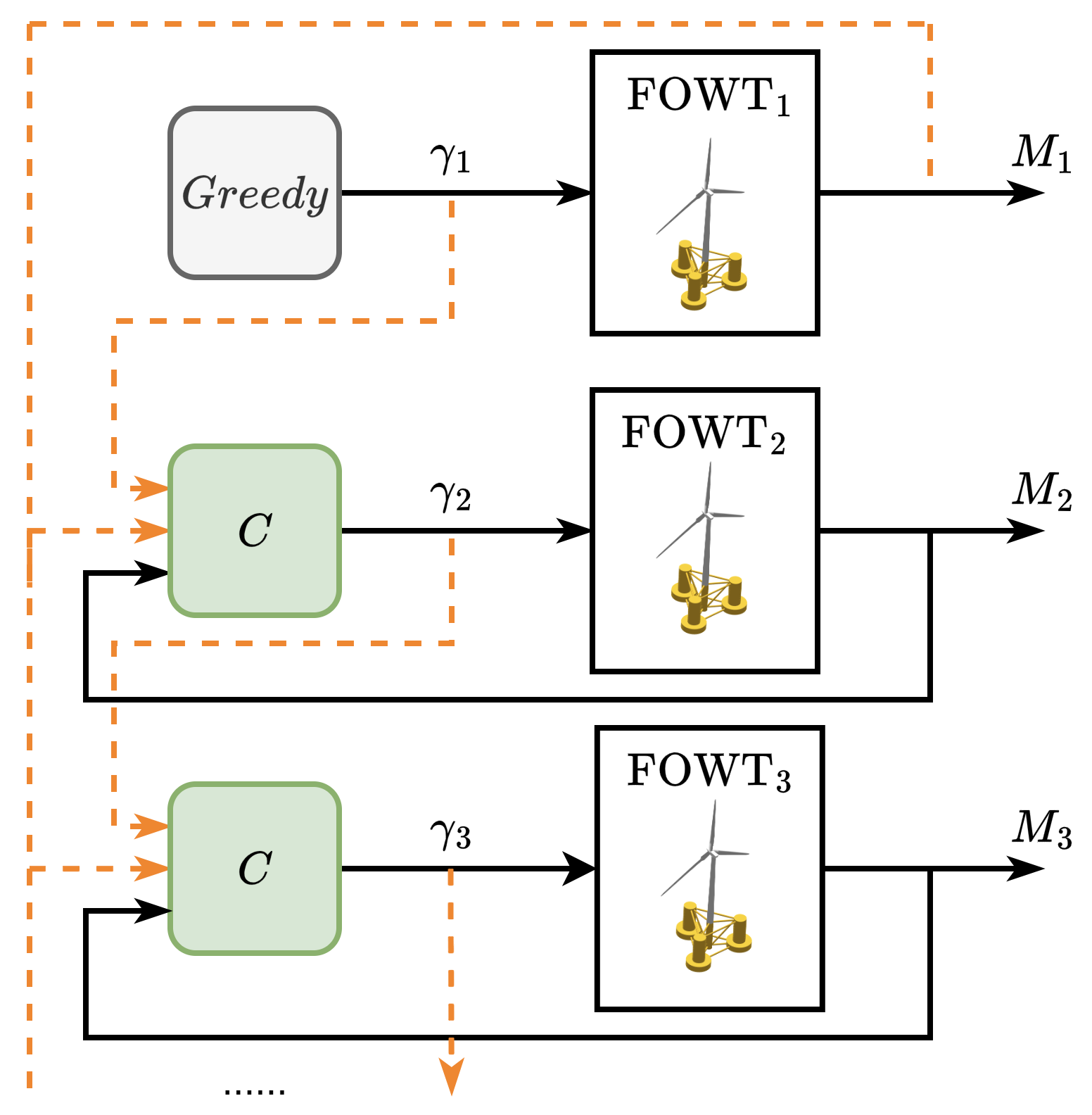}
    \caption{Structure of the proposed control system for a FOWF. Orange lines represent communication among FOWTs.}
    \label{fig:BD}
\end{figure}

Figure~\ref{fig:BD} shows the block diagram of the proposed control structure. For clarity, the time index is omitted, and only three FOWTs are shown. As the figure shows, FOWT\(_1\) operates greedily, consistent with Assumption~\ref{assump:freestream}, and the yellow lines represent communication links, consistent with Assumption~\ref{assump:communication}. The local controller \(C\) determines \(\gamma_{n,t}\) of each FOWT\(_n\) from the broadcast \(M_{1,t}\), the neighbor's \(\gamma_{n-1,t}\), and its own measurement \(M_{n,t}\). =Moreover, Figure~\ref{fig:BD} shows that the system is distributed, as there is no supervisory controller and each non-freestream FOWT makes decisions using the same control block \(C\). The system is also model-free, since no wind farm model is used to derive the optimal yaw angles. Based on the above definitions and assumptions, we formulate the problem of deriving \(\gamma^*_{n,t}\) for each FOWT\(_n\) in the following subsection.

\subsection{Wake Overlap Minimization Problem Formulation}
Turbine repositioning aims to reduce the overlap between the incoming wake and the rotor disk of each FOWT\(_n\). Therefore, we associate a local cost \(f_{n,t}\) with each FOWT\(_n\) that is nonincreasing as the wake overlap of FOWT\(_n\) decreases. Since the wake on FOWT\(_n\) is generated by FOWT\(_{n-1}\), the local cost depends on its own yaw and on that of its neighbor, i.e., \(f_{n,t}(\gamma_{n,t}, \gamma_{n-1,t})\). Its explicit construction from onboard measurements is given in Section~\ref{sec:cost}.

Subsequently, we formulate the control problem under a multi-agent coordination-based distributed online optimization framework~\cite{li2023survey}. At each time step, \(\gamma_{n,t}\) is derived by minimizing \(f_{n,t}(\gamma_{n,t}, \gamma_{n-1,t})\). FOWT\(_n\) then shares its \(\gamma_{n,t}\) and \(M_{1,t}\) with FOWT\(_{n-1}\), and the process repeats. Accordingly, the aggregated wind farm cost is minimized at each time step~\(t\) by solving:
\begin{equation}
\begin{aligned}
    \min_{\{\gamma_{n,t}\}_{n=1}^{N}} \quad
    & f_{1,t}\!\left(\gamma_{1,t}\right) + \sum_{n=2}^{N} f_{n,t}\!\left(\gamma_{n,t},\, \gamma_{n-1,t}\right) \\
    \text{s.t.}\quad
    & \gamma_{1,t} = \gamma_{\text{greedy}} \\
    & |\gamma_{n,t}| \le \gamma_{\max},
\end{aligned}
\label{eq:optiProb}
\end{equation}
\noindent where $\gamma_{\max}$ denotes the maximum nacelle yaw angle imposed to reduce fatigue induced by excessive yawing, and $\gamma_{\text{greedy}}$ denotes the greedy nacelle yaw angle set to track the incoming wind direction.
\section{Distributed Model-Free Control Design} \label{sec:method}
In this section, we propose the methodology.

\subsection{Design Overview}
\begin{figure}[h!]
    \centering
    \includegraphics[width=\linewidth]{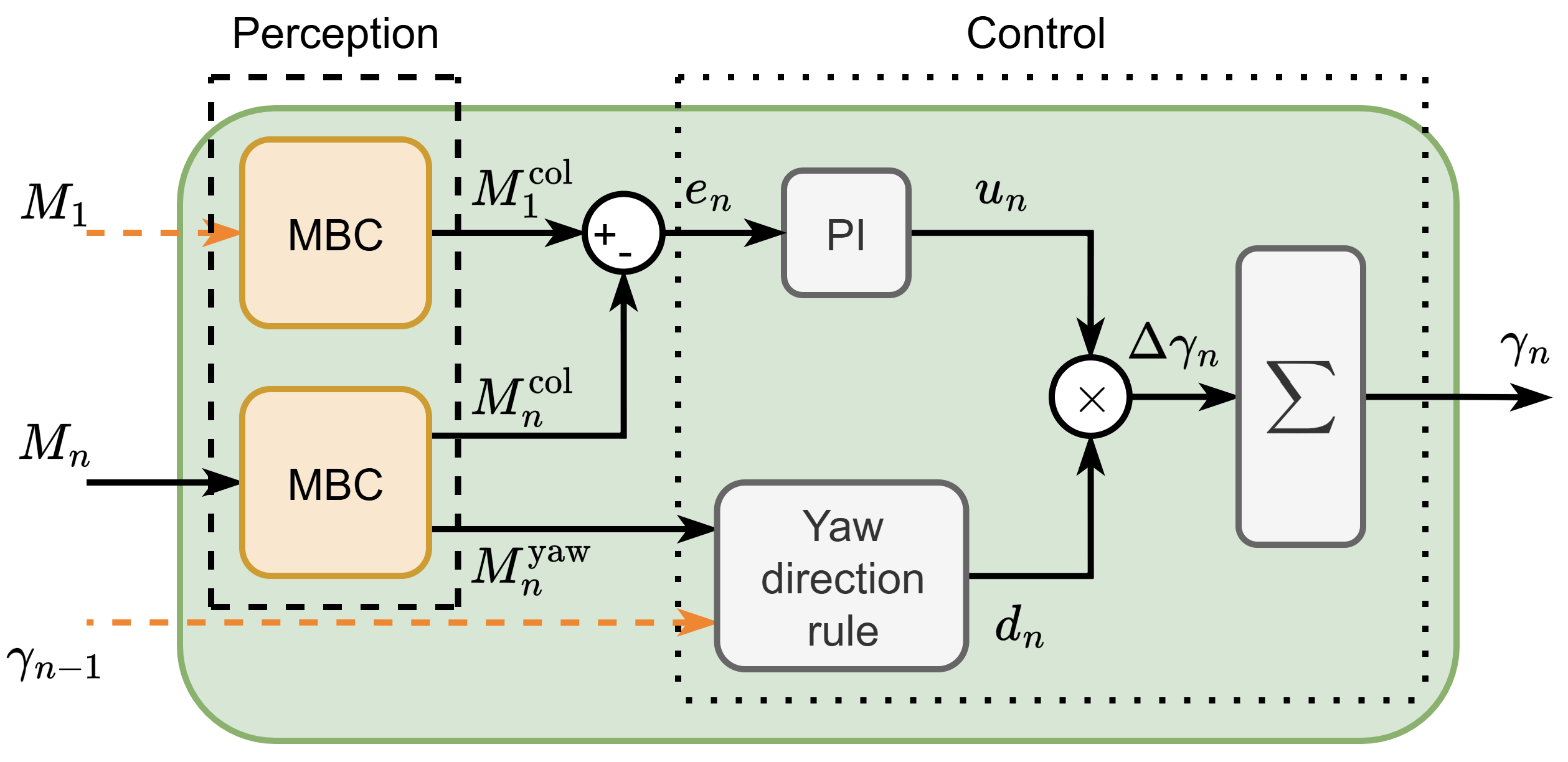}
    \caption{Block diagram of the designed control system $C$. For clarity, the time index is omitted.}
    \label{fig:ctrlSys}
\end{figure}

To solve \eqref{eq:optiProb}, we propose a distributed method that incorporates physics-informed knowledge of wake dynamics for each FOWT\(_n\). To reduce its wake overlap, FOWT\(_n\) must first know how much it is waked and from which direction the wake arrives, and then use this knowledge to reposition itself accordingly. These two steps form a ``perceive-then-control'' principle, in which the control block \(C\) in Figure~\ref{fig:BD} comprises two subsystems. Figure~\ref{fig:ctrlSys} shows the detailed structure of~\(C\), in which the two subsystems are illustrated. The perception subsystem estimates the extent to which FOWT\(_n\) is currently waked and the asymmetric wake distribution over its rotor disk. The control subsystem then determines \(\gamma_{n,t+1}\) to reposition FOWT\(_n\) toward regions of lower wake overlap, using both the estimated wake information and data received from FOWT\(_{n-1}\). The two subsystems are described in detail in Sections~\ref{sec:indirect} and~\ref{sec:derive}.

\subsection{Perception Subsystem: Wake Dynamics Estimation} \label{sec:indirect}
The goal of the perception subsystem is to estimate the wake dynamics of FOWT\(_n\) by determining two features: the extent of wake overlap and the asymmetric wake distribution over its rotor disk. Determining these two features is essential for determining the yaw direction and magnitude, which are discussed later in the paper. 

To ensure practical feasibility, we select \(M_{n,t}\) as the onboard measurement for estimating local wake dynamics, following previous work~\cite{bottasso2018local,bertele2021wind}. The underlying principle is that a higher wind speed produces a larger \(M_{n,t}\). However, \(M_{n,t}\) is measured at the blade root and does not directly represent the rotor disk~\cite{WFFC_Review}. To obtain a stationary representation of the rotor-disk load, we apply the Multi-Blade Coordinate (MBC) transform~\cite{MBC} to map \(M_{n,t}\) into the rotor-disk frame.

\begin{equation}
\label{eq:MBC_equation}
    \begin{bmatrix}
        M^{\mathrm{col}}_{n, t} \\ M^{\mathrm{tilt}}_{n, t} \\ M^{\mathrm{yaw}}_{n, t}
    \end{bmatrix}
    =
    \underbrace{\frac{2}{3} \begin{bmatrix}
        1/2 & 1/2 & 1/2\\
    \cos{\psi^{(1)}_{n,t}} & \cos{\psi^{(2)}_{n,t}} & \cos{\psi^{(3)}_{n,t}} \\
    \sin{\psi^{(1)}_{n,t}} & \sin{\psi^{(2)}_{n,t}} & \sin{\psi^{(3)}_{n,t}} 
    \end{bmatrix}}_\text{MBC Transform Matrix}
    \begin{bmatrix}
        M^{(1)}_{n,t} \\ 
        M^{(2)}_{n,t} \\ 
        M^{(3)}_{n,t}
    \end{bmatrix} \text{.}
\end{equation}

\begin{figure}[h!]
    \centering
    \includegraphics[width=0.5\linewidth]{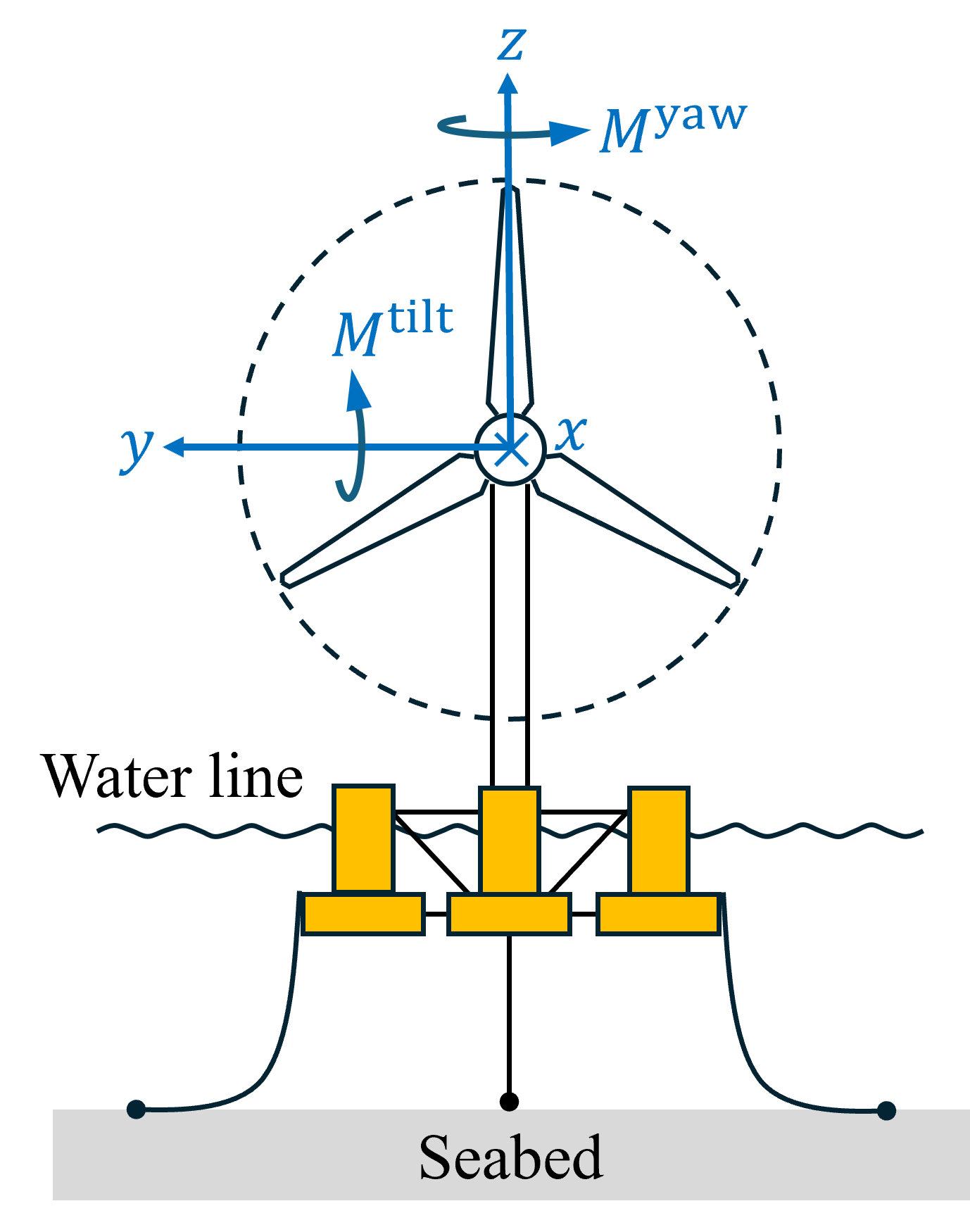}
    \caption{Relationship between \(M^{\mathrm{tilt}}_{n,t}\) and \(M^{\mathrm{yaw}}_{n,t}\) and the rotor-disk load of FOWT\(_n\).}
    \label{fig:MBCtransform}
\end{figure}

Equation~\eqref{eq:MBC_equation} shows the definition of the MBC transform, in which $\psi_{n,t} = \{\psi^{(1)}_{n,t}, \psi^{(2)}_{n,t}, \psi^{(3)}_{n,t}\}$ denotes the azimuth angle of the three blades of FOWT\(_n\), with \(\psi = 0^\circ\) corresponding to the vertically upright position. Specifically, the MBC transform maps the blade loads \(M_{n,t}\) into three equivalent loads in the rotor-disk: the collective moment \(M^{\mathrm{col}}_{n,t}\), the tilt moment \(M^{\mathrm{tilt}}_{n,t}\), and the yaw moment \(M^{\mathrm{yaw}}_{n,t}\). Figure~\ref{fig:MBCtransform} illustrates how \(M^{\mathrm{tilt}}_{n,t}\) and \(M^{\mathrm{yaw}}_{n,t}\) relate to the rotor-disk load, reflecting the vertical and horizontal asymmetry, respectively. Since the out-of-plane bending moment is positively correlated with wind speed, these signals directly reveal the wake distribution over the rotor disk~\cite{bottasso2018local,bertele2021wind}. Specifically, \(M^{\mathrm{col}}_{n,t}\) represents the average wind speed in front of FOWT\(_n\), which gives the extent to which the turbine is waked. In addition, \(M^{\mathrm{yaw}}_{n,t}\) represents the lateral asymmetry of the load, which reveals the relative lateral position of the wake with respect to the rotor disk. Together, these two signals provide the two features required by the perception subsystem for estimating the wake dynamics around FOWT\(_n\).

\begin{figure}[h!]
    \centering
    \includegraphics[width=\linewidth]{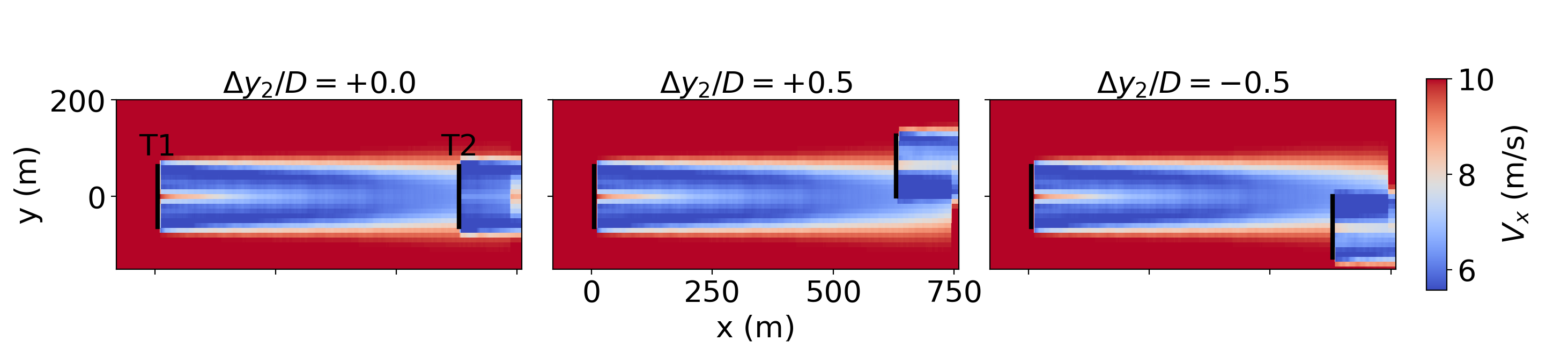}
    \caption{Bird's-eye views of three cases \(\Delta y_2/D = \{\pm0.5, 0\}\). \(V_x\) denotes the streamwise wind speed.}
    \label{fig:MomentWake}
\end{figure}

To validate the above analysis, we simulate a \(1\times2\) FOWF with two NREL 5~MW FOWTs~\cite{NREL5MW,NREL5MW_FOWT} spaced \(5D\) apart, under steady \(10\)~m/s hub-height wind in FAST.Farm~\cite{FastFarm}. Figure~\ref{fig:MomentWake} shows this setup, in which FOWT\(_2\) is placed at different lateral crosswind positions relative to FOWT\(_1\), quantified by \(\Delta y_2/D\), where \(y_2\) denotes the lateral position of FOWT\(_2\).

\begin{figure}[h!]
    \centering
    \includegraphics[width=\linewidth]{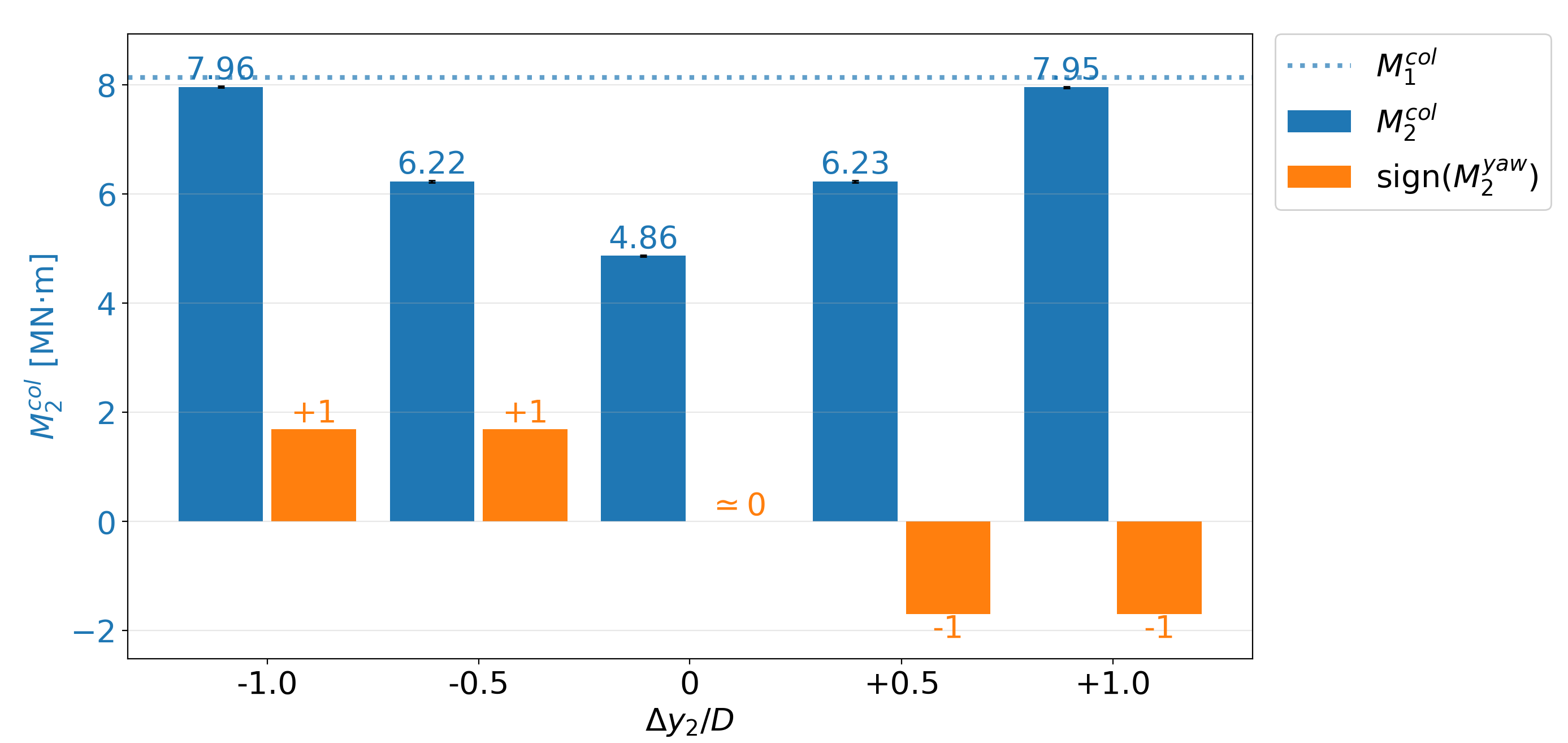}
    \caption{Steady-state values of \(M^{\mathrm{col}}_2\) and \(\mathrm{sign}(M^{\mathrm{yaw}}_2)\) for \(\Delta y_2/D = \{\pm1, \pm0.5, 0\}\).}
    \label{fig:MomentWake2}
\end{figure}

Figure~\ref{fig:MomentWake2} shows the steady-state values of \(M^{\mathrm{col}}_2\) and \(\mathrm{sign}(M^{\mathrm{yaw}}_2)\) for \(\Delta y_2/D = \{\pm1, \pm0.5, 0\}\), from which two indicators for wake overlap information can be derived:
\subsubsection{Indicator of Wake Overlap Extent}
As FOWT\(_2\) moves farther out of the wake and experiences less wake overlap, \(M^{\mathrm{col}}_2\) increases and approaches the freestream value \(M^{\mathrm{col}}_1\). This trend is independent of whether FOWT\(_2\) is to the left or the right of FOWT\(_1\).

\subsubsection{Indicator of Lateral Wake Asymmetry} The sign of \(M^{\mathrm{yaw}}_2\) changes depending on the lateral side of FOWT\(_2\) relative to FOWT\(_1\), thereby reflecting the lateral asymmetry of the wake over the rotor disk.

Therefore, these findings are consistent with the earlier analysis, which shows that the combination of \(M_{n,t}\) and the MBC transform enables indirect estimation of two features that are essential for determining \(\gamma_{n,t+1}\).

\subsection{Control Subsystem} \label{sec:derive}
In this section, we design the control subsystem that derives \(\gamma_{n,t+1}\) for each FOWT\(_n\). We first define the cost function \(f_{n,t}\) and then design the algorithm to derive \(\gamma_{n,t+1}\). 

\subsubsection{Cost Function Design} \label{sec:cost}
Based on the relationship established in Section~\ref{sec:indirect}, we define \(f_{n,t}\) to link the control objective of minimizing wake overlap directly to the wake dynamics through \(M^{\mathrm{col}}_{n,t}\):
\begin{equation} \label{eq:cost}
    f_{n,t} \coloneq \frac{M^{\mathrm{col}}_{1,t} - M^{\mathrm{col}}_{n,t}}{M^{\mathrm{col}}_{1,t}}.
\end{equation}

Equation~\eqref{eq:cost} defines \(f_{n,t}\) as the normalized deficit of \(M^{\mathrm{col}}_{n,t}\) relative to~\(M^{\mathrm{col}}_{1,t}\). This definition uses FOWT\(_1\) as the reference because it represents the best attainable performance under the given wind condition. Therefore, minimizing \(f_{n,t}\) drives each FOWT\(_n\) toward the performance of FOWT\(_1\). As established in Section~\ref{sec:indirect}, \(M^{\mathrm{col}}_{n,t}\) is influenced by \(\gamma_{n,t}\) and \(\gamma_{n-1,t}\), and \(f_{n,t}\) decreases as the wake overlap of FOWT\(_n\) decreases, which is consistent with the control objective. Excessive yawing is also penalized, since yawing reduces \(M^{\mathrm{col}}_{n,t}\) and thus increases the cost~\cite{thedin2025load}.

\subsubsection{Control Algorithm} \label{sec:algorithm}
We now design an algorithm to derive \(\gamma_{n,t+1}\) by minimizing \(f_{n,t}\). As established in Section~\ref{sec:indirect}, the perception subsystem provides the extent of wake overlap and the asymmetric wake distribution over the rotor disk of FOWT\(_n\). Subsequently, we use these features to determine the yaw increment~\(u_{n,t+1}\) and a yaw direction~\(d_{n,t}\), which combine to~\(\Delta\gamma_{n,t+1}\).

Specifically, \(u_{n,t}\) is determined using~$M^{\mathrm{col}}$. Since minimizing \(f_{n,t}\) is equivalent to minimizing the difference \(e_{n,t} \coloneq M^{\mathrm{col}}_{1,t} - M^{\mathrm{col}}_{n,t}\), a simple Proportional-Integral (PI) controller with \(K_p = 0.8\) and \(K_i = 0.0125\) is used to determine~$u_{n,t+1}$. Equation~\eqref{eq:pid} shows the output of the incremental PI form:
\begin{equation}
    \label{eq:pid}
    u_{n,t} = K_p \left(e_{n,t} - e_{n,t-1}\right) + K_i \, e_{n,t} \, \Delta t \text{.}
\end{equation}

The yaw direction \(d_{n,t}\) is determined by a rule block that takes \(M^{\mathrm{yaw}}_{n,t}\) and \(\gamma_{n-1,t}\) as inputs and Equation~\eqref{eq:direction} shows the implemented logic. The principle is to direct FOWT\(_n\) toward regions with less wake while avoiding a direction decision that would conflict with FOWT\(_{n-1}\) and steer FOWT\(_n\) into its wake:

\begin{equation} \label{eq:direction}
    d_{n,t} =
    \begin{cases}
        -\operatorname{sign}\!\left(M^{\mathrm{yaw}}_{n,t}\right), & |\gamma_{n-1,t}| \leq \epsilon, \\[6pt]
        -\operatorname{sign}\!\left(\gamma_{n-1,t}\right), & \text{otherwise},
    \end{cases}
\end{equation}
where \(\epsilon\) is a small threshold. When the upstream turbine is effectively stationary, \(d_{n,t}\) is determined from \(M^{\mathrm{yaw}}_{n,t}\). Otherwise, \(d_{n,t}\) takes the opposite direction of \(\gamma_{n-1,t}\) to avoid a conflicting decision. 

In summary, \(d_{n,t}\) fixes the side toward which FOWT\(_n\) repositions, while the PI output controls the size of the increment, and the two signals combine.
\section{CASE STUDY} \label{sec:case}
In this section, we demonstrate the feasibility and the effectiveness of the proposed method through a case study.

\subsection{Simulation Setup}
As a proof of concept, the proposed method is validated on a simple \(1\times3\) array FOWF composed of NREL 5~MW semi-submersible FOWTs~\cite{NREL5MW, NREL5MW_FOWT}. The spacing between adjacent FOWTs is \(5D\) with no lateral offset. To enable turbine repositioning, the mooring line length is extended to 970 m~\cite{kheirabadi2019modeling}. A steady inflow condition is used, with the hub-height wind speed held constant at \(10\)~m/s throughout the simulation. The maximum yaw angle is set to \(\gamma_{\max}=30^\circ\) following industry practice~\cite{zong2021experimental}, and the sampling rate of all signals is set to \(10\) Hz.

All simulations were run for \(3000\)~s in FAST.Farm~\cite{FastFarm} using a self-developed interface~\cite{ffconnect} on a laptop equipped with a 24-core Intel Core Ultra CPU and 32 GB of RAM.

\subsection{Performance Metrics}
To assess the performance of the proposed method, we use three metrics. The first metric is the aggregated cost of all FOWTs, which quantifies the extent of wake overlap in the FOWF. The second metric is the farm-level power output. The third metric is the static regret \(S\text{-}Reg\), defined as:
\begin{equation} \label{eq:SReg}
    S\text{-}Reg \coloneqq \sum_{t=1}^{T} \bigl( f_t(\Gamma_t) - f_t(\Gamma^*) \bigr),
\end{equation}
where \(f_t(\Gamma_t) \coloneqq f_{1,t}(\gamma_{1,t}) + \sum_{n=2}^{N} f_{n,t}(\gamma_{n,t}, \gamma_{n-1,t})\) is the farm-level cost for a given yaw vector \(\Gamma_t\), and~\(\Gamma^* \coloneqq (\gamma_{1}^*, \dots, \gamma_{N}^*)\) denotes the vector of optimal yaw angles for the wind farm. Thus, \(S\text{-}Reg\) compares the total cost incurred by the online algorithm with the minimal cost attainable under a fixed yaw vector held constant over the entire time horizon. The performance of an online algorithm is considered good if its $ S\text{-}Reg$ is bounded by a sublinear function of \(T\)~\cite{li2023survey}. For comparison, \(\Gamma^*\) is set to the yaw angles obtained from the baseline model-based optimization~\cite{kheirabadi2020real}, which serve as a surrogate for the true \(\Gamma^*\).

\begin{figure}[h!]
    \centering
    \begin{subfigure}{\linewidth}
        \centering
        \includegraphics[width=\linewidth]{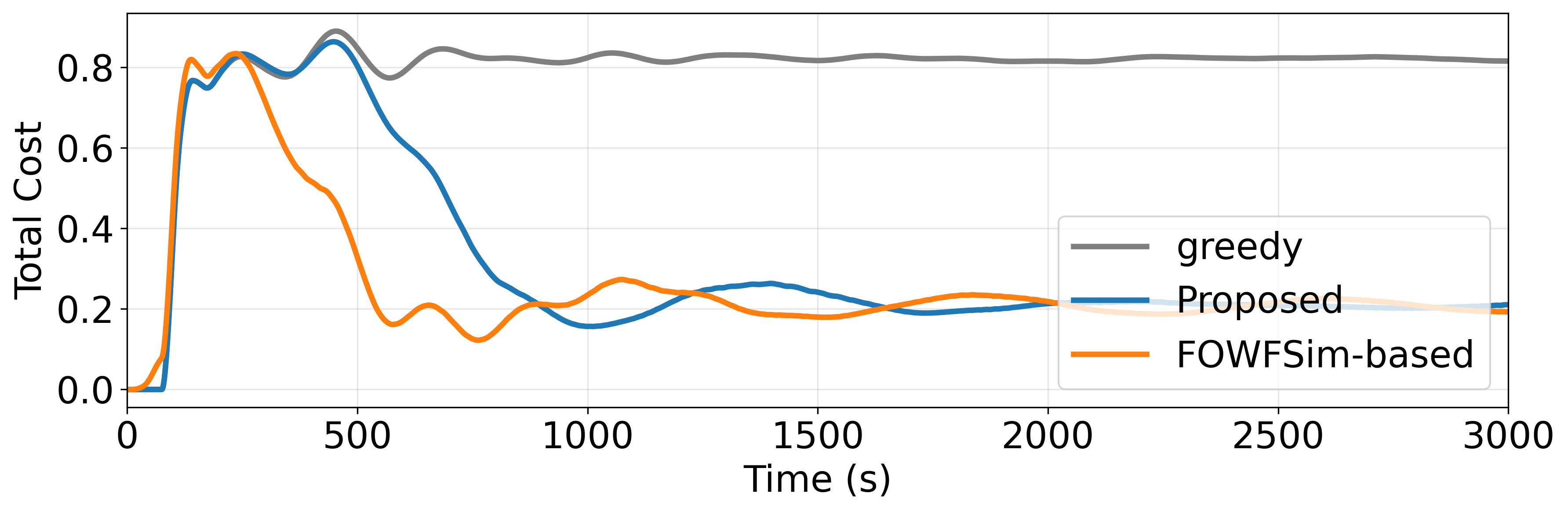}
        \caption{Aggregated wind farm cost.}
        \label{fig:ts_cost}
    \end{subfigure}\\[4pt]
    \begin{subfigure}{\linewidth}
        \centering
        \includegraphics[width=\linewidth]{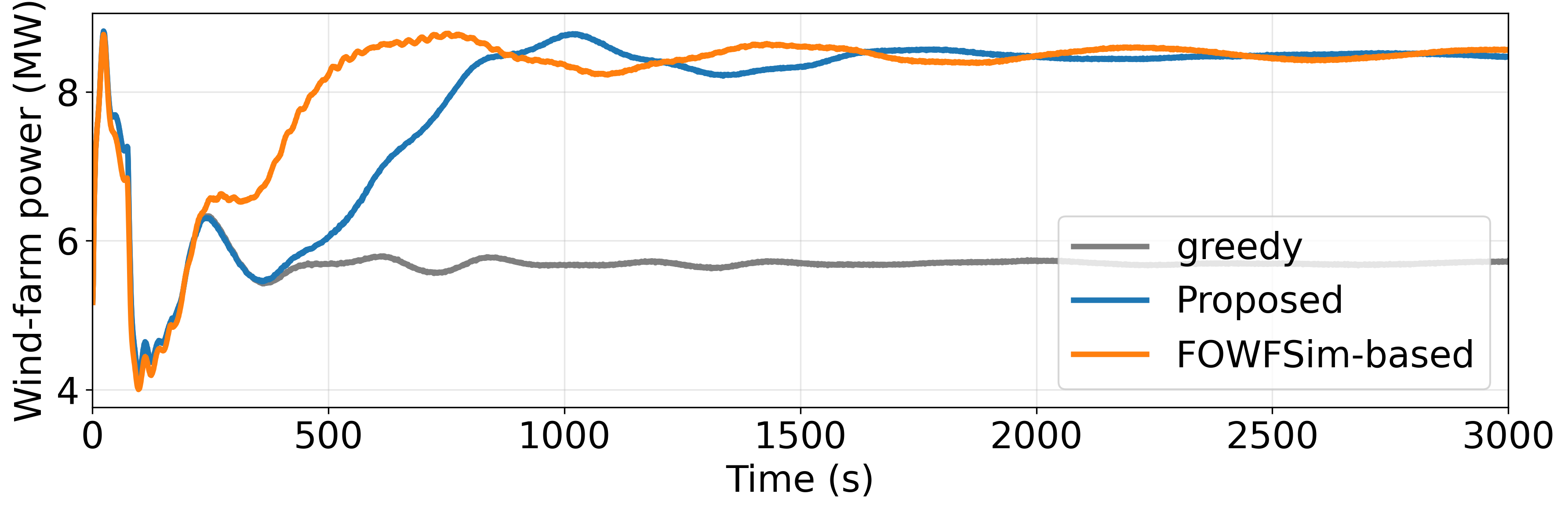}
        \caption{Aggregated wind farm power output.}
        \label{fig:ts_power}
    \end{subfigure}\\[4pt]
    \begin{subfigure}{\linewidth}
        \centering
        \includegraphics[width=\linewidth]{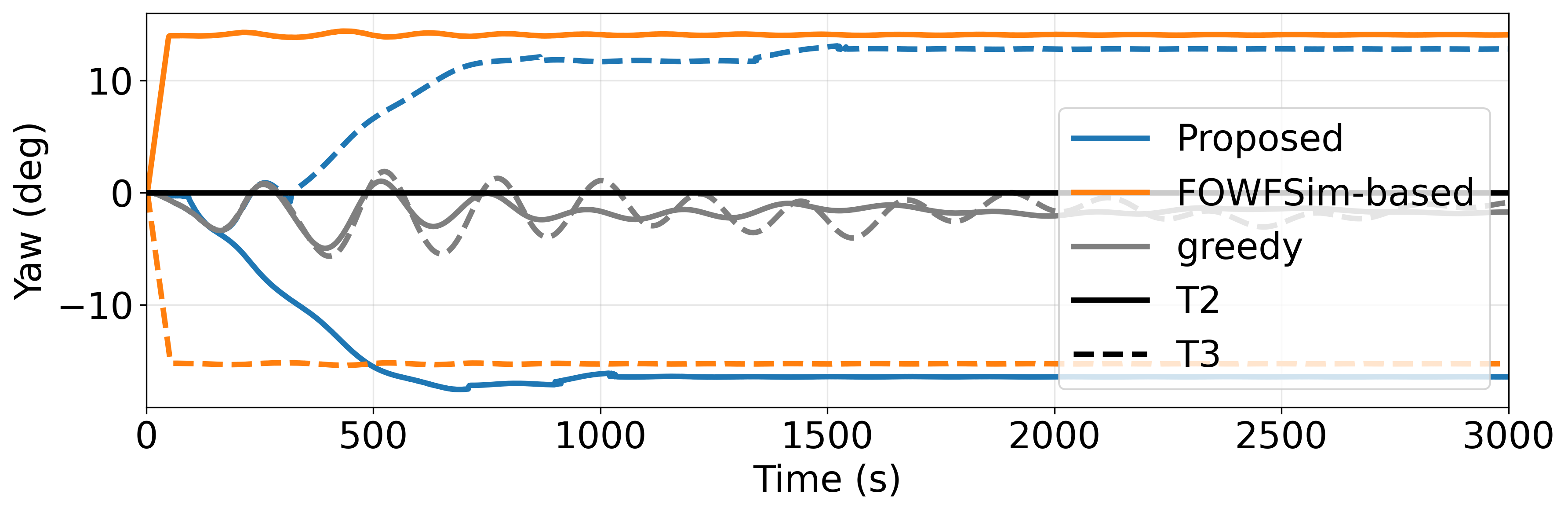}
        \caption{Optimal yaw angles.}
        \label{fig:ts_yaw}
    \end{subfigure}\\[4pt]
    \begin{subfigure}{\linewidth}
        \centering
        \includegraphics[width=\linewidth]{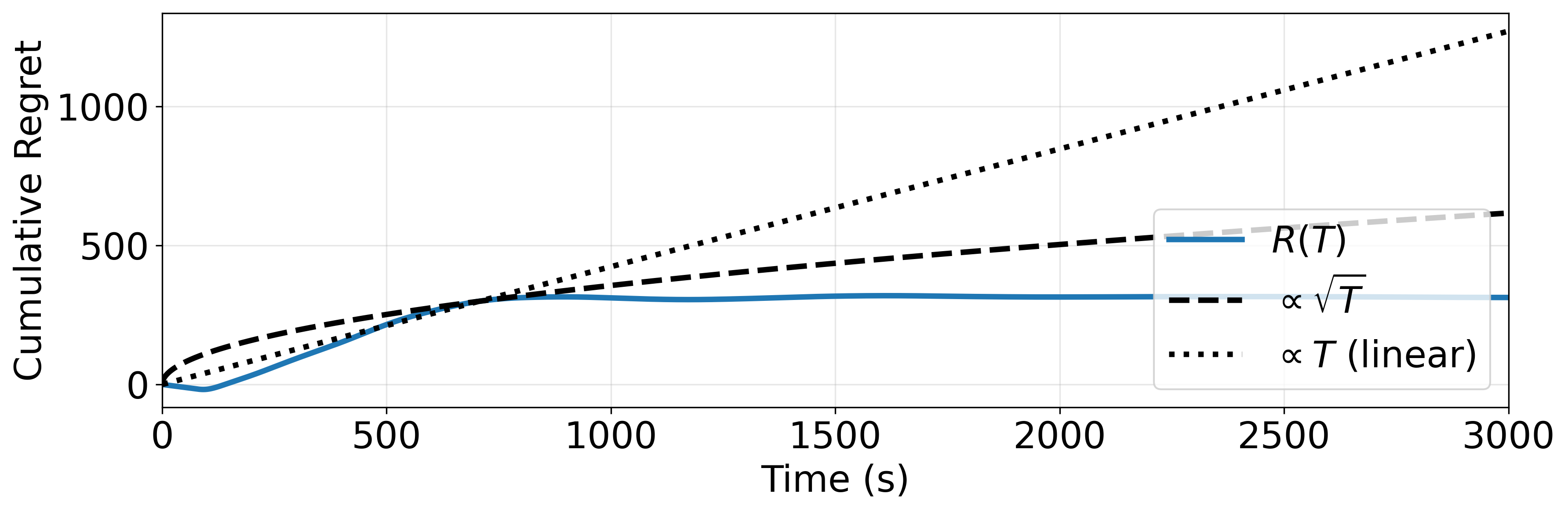}
        \caption{Static regret $S\text{-}Reg$ versus sublinear bounds.}
        \label{fig:ts_regret}
    \end{subfigure}
    \caption{Time-varying simulation results for the greedy (grey), FOWFSim-based (yellow), and proposed (blue) cases: (a) cost, (b) power, (c) yaw actions, and (d) static regret.}
    \label{fig:timeseries}
\end{figure}

\subsection{Simulation Results}
In this section, we show the simulation results. To assess performance, the proposed method is compared with a greedy baseline, in which no wind farm control is applied and each FOWT simply runs a baseline controller, and with a model-based optimization baseline that uses the low-fidelity wind farm model FOWFSim to derive \(\gamma^*_{n,t}\)~\cite{kheirabadi2020real}.

\subsubsection{Total Cost}
Figure~\ref{fig:timeseries}(a) shows the evolution of the farm-level cost over time. The cost decreases and converges after a transition period of approximately 1000~s. Compared with the model-based case, the proposed model-free method exhibits a slower response, which is consistent with the behavior reported in existing work~\cite{kheirabadi2019quantitative}. Eventually, the cost converges to a value comparable to that of the model-based case, showing that the proposed method is able to mitigate the wake effect.

\subsubsection{Power Production}
Figure~\ref{fig:timeseries}(b) further shows the aggregated wind farm power output throughout the simulation. In steady state, the FOWFSim-based case reaches $8.488$~MW and the proposed distributed case reaches $8.499$~MW, corresponding to increases of $49.1\%$ and $49.3\%$, respectively, relative to the greedy baseline of $5.693$ MW. These results demonstrate the feasibility of the proposed method for improving FOWF performance.

\subsubsection{Yaw Actuation}
The yaw actions \(\gamma_{n,t}\) derived by both the FOWFSim-based and the proposed methods are shown in Figure~\ref{fig:timeseries}(c). As expected, the proposed distributed case converges more slowly to \(\gamma_{n,t}\) than the FOWFSim-based case. Nevertheless, the resulting \(\gamma_{n,t}\) values for FOWT\(_2\) and FOWT\(_3\) are of comparable magnitude, which confirms that the cost function effectively balances the trade-off between minimizing wake overlap and avoiding excessive yawing.

\subsubsection{Static Regret}
Figure~\ref{fig:timeseries}(d) compares $S\text{-}Reg$ with a family of sublinear reference functions of $T$ with scaling factors. As shown, $S\text{-}Reg$ remains below all considered sublinear bounds, which validates the performance of the proposed distributed online optimization-based method.

\subsubsection{Steady-State Wake Field}
\begin{figure}[h!]
    \centering
    \includegraphics[width=\linewidth]{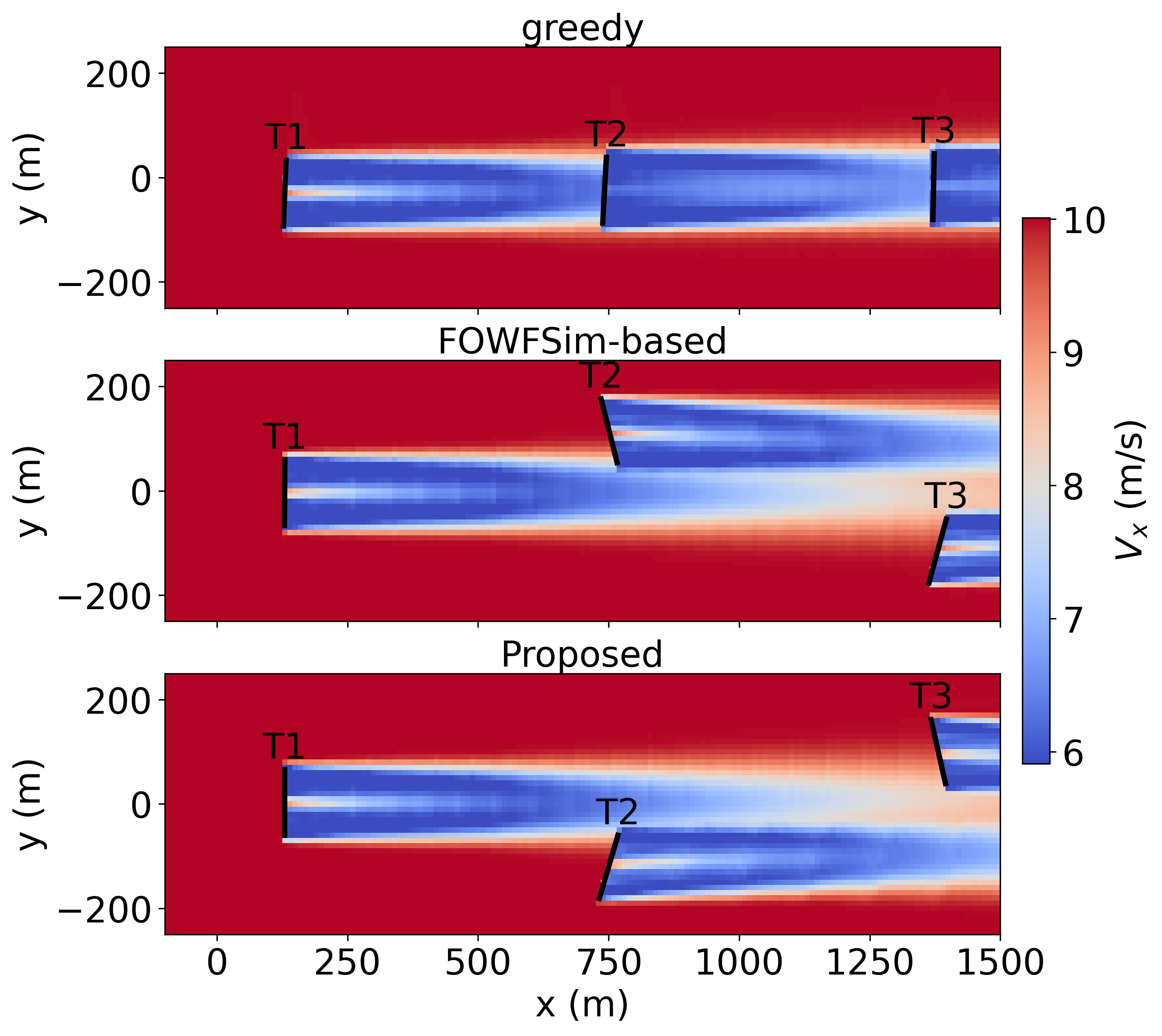}
    \caption{Comparison of the steady-state wake fields from FAST.Farm for the three cases.}
    \label{fig:snapshotCompare}
\end{figure}

The instantaneous wake field of the FOWF is shown in Figure~\ref{fig:snapshotCompare}. The blue region represents the wake, which downstream turbines FOWT\(_2\) and FOWT\(_3\) seek to avoid. Both turbines yaw toward the red region, where the wind speed is higher. Overall, both the proposed method and the FOWFSim-based method mitigate the wake effect by minimizing the wake overlap of downstream turbines compared to the greedy case.

\subsubsection{Computation Time}
Finally, we compare the computation times of the FOWFSim-based and proposed cases. To ensure a fair comparison, only the time from acquiring the measurements to sending the $\gamma_{n, t+1}$ is considered. On average, the FOWFSim-based method requires \(0.2313\)~s per iteration, whereas the proposed method requires only \(0.0746\)~ms. This corresponds to a more than 3000-fold reduction in computation time, which results from the distributed formulation of the control problem and the model-free local logic that avoids gradient computation.

Overall, the simulation results indicate that the proposed method effectively mitigates the wake effect while remaining efficient enough for real-time operation.

\section{CONCLUSION} \label{sec:conclusion}
In this paper, we introduced a model-free distributed method for turbine repositioning control with the goal of minimizing wake overlap for floating offshore wind farms. We first formulated the problem of deriving the optimal yaw angle within a multi-agent distributed online optimization framework and proposed a model-free method to solve it. To address the nonconvexity and inaccessible gradient information of the problem, the proposed method adjusts the optimal yaw angles based on physics-informed knowledge of wake dynamics, which are estimated from the onboard blade load signals to ensure practical feasibility. We further designed the cost function based on the estimated wake dynamics to indicate the degree of wake overlap in the wind farm. The proposed method was shown to be effective in the case study, achieving a significant reduction in computational effort relative to the model-based optimization approach while attaining slightly better power output. Future work will extend the communication and control schemes to account for more complex aerodynamic interactions in non-array FOWF configurations and to handle time-varying wind conditions.

% \addtolength{\textheight}{-12cm}   % formating

%%%%%%%%%%%%%%%%%%%%%%%%%%%%%%%%%%%%%%%%%%%%%%%%%%%%%%%%%%%%%%%%%%%%%%%%%%%%%%%%
% \section*{ACKNOWLEDGMENT}
% Generative AI is used for polishing writing.

%%%%%%%%%%%%%%%%%%%%%%%%%%%%%%%%%%%%%%%%%%%%%%%%%%%%%%%%%%%%%%%%%%%%%%%%%%%%%%%%
\bibliographystyle{IEEEtran}
\bibliography{ref}

@article{zong2021experimental,
  title={Experimental investigation and analytical modelling of active yaw control for wind farm power optimization},
  author={Zong, Haohua and Port{\'e}-Agel, Fernando},
  journal={Renewable Energy},
  doi={10.1016/j.renene.2021.02.059},
  volume={170},
  pages={1228--1244},
  year={2021},
  publisher={Elsevier}
}

@book{busby2012wind,
  title={Wind Power: The Industry Grows Up},
  author={Busby, Rebecca L},
  year={2012},
  publisher={PennWell Corp.},
  isbn={978-1-59370-244-1},
}

@article{WFFC_Review,
	title = {Wind farm flow control: prospects and challenges},
	volume = {7},
	issn = {2366-7443},
	shorttitle = {Wind farm flow control},
	doi = {10.5194/wes-7-2271-2022},
	number = {6},
	journal = {Wind Energy Science},
	author = {Meyers, Johan and Bottasso, Carlo and Dykes, Katherine and Fleming, Paul and Gebraad, Pieter and Giebel, Gregor and Göçmen, Tuhfe and van Wingerden, Jan-Willem},
	month = nov,
	year = {2022},
	pages = {2271--2306},
}

@article{kheirabadi2019quantitative,
  title={A quantitative review of wind farm control with the objective of wind farm power maximization},
  author={Kheirabadi, Ali C and Nagamune, Ryozo},
  journal={Journal of Wind Engineering and Industrial Aerodynamics},
  doi={10.1016/j.jweia.2019.06.015},
  volume={192},
  pages={45--73},
  year={2019},
  publisher={Elsevier}
}

@article{niu2024floating,
  title={Floating offshore wind farm control via turbine repositioning: Unlocking the potential unique to floating offshore wind},
  author={Niu, Yue and Dwivedi, Arpit and Sathiaraj, Joel and Lathi, Parth P and Nagamune, Ryozo},
  journal={IEEE Control Systems Magazine},
  doi={10.1109/MCS.2024.3432342},
  volume={44},
  number={5},
  pages={106--129},
  year={2024},
  publisher={IEEE}
}

@inproceedings{kheirabadi2019modeling,
  title={Modeling and power optimization of floating offshore wind farms with yaw and induction-based turbine repositioning},
  author={Kheirabadi, Ali C and Nagamune, Ryozo},
  doi={10.23919/ACC.2019.8814600},
  booktitle={2019 American Control Conference (ACC)},
  pages={5458--5463},
  year={2019},
  organization={IEEE}
}

@article{kheirabadi2020real,
  title={Real-time relocation of floating offshore wind turbine platforms for wind farm efficiency maximization: An assessment of feasibility and steady-state potential},
  author={Kheirabadi, Ali C and Nagamune, Ryozo},
  journal={Ocean Engineering},
  volume={208},
  pages={107445},
  year={2020},
  publisher={Elsevier}
}

@article{wei2025multi,
  title={Multi-object reposition control of floating wind farms considering time-varying change of wind},
  author={Wei, Shangshang and Li, Zhihan and Wang, Xin and Feng, Dachuan and Gao, Xianhua},
  journal={Ocean Engineering},
  volume={327},
  pages={120974},
  year={2025},
  publisher={Elsevier}
}

@inproceedings{niu2025power,
  title={Power Maximization and Platform Oscillation Mitigation in Reconfigurable Floating Offshore Wind Farms},
  author={Niu, Yue and Nagamune, Ryozo},
  booktitle={2025 American Control Conference (ACC)},
  pages={1488--1493},
  year={2025},
  organization={IEEE}
}

@article{huang2025bi,
  title={Bi-level multi-objective optimization framework for wake escape in floating offshore wind farm},
  author={Huang, Chaoneng and Wang, Li and Huang, Qian and Song, Dongran and Yang, Jian and Dong, Mi and Joo, Young Hoo and Dui{\'c}, Neven},
  journal={Applied Energy},
  volume={377},
  pages={124712},
  year={2025},
  publisher={Elsevier}
}

@article{jard2025real,
  title={Real-time dynamic layout optimization for floating offshore wind farm control},
  author={Jard, Timoth{\'e} and Snaiki, Reda},
  journal={Ocean Engineering},
  volume={316},
  pages={119971},
  year={2025},
  publisher={Elsevier}
}

@inproceedings{boersma2017tutorial,
  title={A tutorial on control-oriented modeling and control of wind farms},
  author={Boersma, Sjoerd and Doekemeijer, Bart M and Gebraad, Pieter MO and Fleming, Paul A and Annoni, Jennifer and Scholbrock, Andrew K and Frederik, Joeri Alexis and van Wingerden, Jan-Willem},
  booktitle={2017 American control conference (ACC)},
  pages={1--18},
  year={2017},
  organization={IEEE}
}

@article{mei2025improving,
  title={Improving floating offshore wind farm flow control with scalable model-based deep reinforcement learning},
  author={Mei, Mingyang and Kou, Peng and Xu, Yilin and Zhang, Zhihao and Tian, Runze and Liang, Deliang},
  journal={IEEE Transactions on Automation Science and Engineering},
  doi={10.1109/TASE.2025.3585016},
  volume={22},
  pages={18255--18268},
  year={2025},
  publisher={IEEE}
}

@article{bottasso2018local,
  title={Local wind speed estimation, with application to wake impingement detection},
  author={Bottasso, CL and Cacciola, S and Schreiber, J},
  journal={Renewable Energy},
  doi={10.1016/j.renene.2017.09.044},
  volume={116},
  pages={155--168},
  year={2018},
  publisher={Elsevier}
}

@article{thedin2025load,
  title={Load assessment of a wind farm considering negative and positive yaw misalignment for wake steering},
  author={Thedin, Regis and Barter, Garrett and Jonkman, Jason and Mudafort, Rafael and Bay, Christopher J and Shaler, Kelsey and Kreeft, Jasper},
  journal={Wind Energy Science},
  doi={10.5194/wes-10-1033-2025},
  volume={10},
  number={6},
  pages={1033--1053},
  year={2025},
  publisher={Copernicus Publications G{\"o}ttingen, Germany}
}

@article{bertele2021wind,
  title={Wind inflow observation from load harmonics: initial steps towards a field validation},
  author={Bertel{\`e}, Marta and Bottasso, Carlo L and Schreiber, Johannes},
  journal={Wind Energy Science},
  doi={10.5194/wes-6-759-2021},
  volume={6},
  number={3},
  pages={759--775},
  year={2021},
  publisher={Copernicus Publications G{\"o}ttingen, Germany}
}

@article{marden2013model,
  title={A model-free approach to wind farm control using game theoretic methods},
  author={Marden, Jason R and Ruben, Shalom D and Pao, Lucy Y},
  journal={IEEE Transactions on Control Systems Technology},
  doi={10.1109/TCST.2013.2257780},
  volume={21},
  number={4},
  pages={1207--1214},
  year={2013},
  publisher={IEEE}
}

@book{burton2011wind,
  title={Wind Energy Handbook},
  author={Burton, Tony and Jenkins, Nick and Sharpe, David and Bossanyi, Ervin},
  year={2011},
  publisher={John Wiley \& Sons},
  isbn = {9780470699751},
}

@article{li2023survey,
  title={A survey on distributed online optimization and online games},
  author={Li, Xiuxian and Xie, Lihua and Li, Na},
  journal={Annual Reviews in Control},
  doi={10.1016/j.arcontrol.2023.100904.},
  volume={56},
  pages={100904},
  year={2023},
  publisher={Elsevier}
}

@inproceedings{MBC,
  title={Multi-blade coordinate transformation and its application to wind turbine analysis},
  author={Bir, Gunjit},
  booktitle={46th AIAA aerospace sciences meeting and exhibit},
  doi={10.2514/6.2008-1300},
  pages={1300},
  year={2008}
}

@book{FastFarm,
  title={{FAST.Farm} user's guide and theory manual},
  url={https://docs.nlr.gov/docs/fy21osti/78485.pdf},
  author={Jonkman, Jason Mark and Shaler, Kelsey},
  year={2021},
  publisher={National Renewable Energy Laboratory, Golden, CO, USA}
}

@techreport{NREL5MW_FOWT,
  title={Definition of the semisubmersible floating system for phase {II} of {OC4}},
  url={https://docs.nlr.gov/docs/fy14osti/60601.pdf},
  author={Robertson, Amy and Jonkman, Jason and Masciola, Marco and Song, Huimin and Goupee, Andrew and Coulling, Alexander and Luan, C},
  year={2014},
  institution={National Renewable Energy Laboratory (NREL), Golden, CO.}
}

@misc{ffconnect,
      title={An Interactive Interface for Control Integration in Mid-Fidelity Wind Farm Simulation}, 
      author={Zekai Chen and Casey Heiskell and Ryozo Nagamune},
      year={2026},
      eprint={2607.16386},
      archivePrefix={arXiv},
      primaryClass={eess.SY},
      url={https://arxiv.org/abs/2607.16386}, 
}

@techreport{NREL5MW,
  title={Definition of a 5-{MW} reference wind turbine for offshore system development},
  author={Jonkman, Jason and Butterfield, Sandy and Musial, Walter and Scott, George},
  year={2009},
  institution={National Renewable Energy Laboratory (NREL), Golden, CO.}
}

\end{document}